\documentclass[twocolumn,amsmath,showpacs,nofootinbib]{revtex4-1}
\usepackage{graphicx}
\usepackage{dcolumn}
\usepackage{bm}
\usepackage{color} 
\usepackage{slashed}
\begin{document}
%\preprint{APS/123-QED}
%%%%%%%%%%%%%%%%%%%%%%%%
\newcommand{\hs}{\hspace*{0.5cm}}
\newcommand{\vs}{\vspace*{0.5cm}}
\newcommand{\be}{\begin{equation}}
\newcommand{\ee}{\end{equation}}
\newcommand{\bea}{\begin{eqnarray}}
\newcommand{\eea}{\end{eqnarray}}
\newcommand{\ben}{\begin{enumerate}}
\newcommand{\een}{\end{enumerate}}
\newcommand{\bde}{\begin{widetext}}
\newcommand{\ede}{\end{widetext}}
\newcommand{\nn}{\nonumber}
\newcommand{\crn}{\nonumber \\}
\newcommand{\Tr}{\mathrm{Tr}}
\newcommand{\non}{\nonumber}
\newcommand{\noi}{\noindent}
\newcommand{\al}{\alpha}
\newcommand{\la}{\lambda}
\newcommand{\bet}{\beta}
\newcommand{\ga}{\gamma}
\newcommand{\va}{\varphi}
\newcommand{\om}{\omega}
\newcommand{\pa}{\partial}
\newcommand{\+}{\dagger}
\newcommand{\fr}{\frac}
\newcommand{\bc}{\begin{center}}
\newcommand{\ec}{\end{center}}
\newcommand{\Ga}{\Gamma}
\newcommand{\de}{\delta}
\newcommand{\De}{\Delta}
\newcommand{\ep}{\epsilon}
\newcommand{\varep}{\varepsilon}
\newcommand{\ka}{\kappa}
\newcommand{\La}{\Lambda}
\newcommand{\si}{\sigma}
\newcommand{\Si}{\Sigma}
\newcommand{\ta}{\tau}
\newcommand{\up}{\upsilon}
\newcommand{\Up}{\Upsilon}
\newcommand{\ze}{\zeta}
\newcommand{\ps}{\psi}
\newcommand{\Ps}{\Psi}
\newcommand{\ph}{\phi}
\newcommand{\vph}{\varphi}
\newcommand{\Ph}{\Phi}
\newcommand{\Om}{\Omega}
\newcommand{\AdrHEPC}{Phenikaa Institute for Advanced Study and Faculty of Basic Science, Phenikaa University, Yen Nghia, Ha Dong, Hanoi 100000, Vietnam}
%%%%%%%%%%%%%%%%%%%%%%%%

\title{Physics implication from a $Z_3$ symmetry of matter} 

\author{Phung Van Dong} 
\email{dong.phungvan@phenikaa-uni.edu.vn}
\affiliation{\AdrHEPC} 

\date{\today}

\begin{abstract}

I show that breaking $B-L$ by one unit of this charge is suitable for neutrino mass generation through an inverse seesaw mechanism, stabilizing a dark matter candidate without supersymmetry, as well as solving the muon anomalous magnetic moment and the $W$ mass deviation via dark field contributions. The new physics is governed by the residual $Z_3$ symmetry of $B-L$ isomorphic to the center of the color group, instead of the well-studied matter parity.     
                  
\end{abstract} 

\maketitle

{\it Introduction}.---Of the exact conservations in physics, the conservation of baryon number minus lepton number, say $B-L$, is questionable. Of the fundamental dynamics in physics, the confinement of colors within hadrons that allows only hadronic states of types $qqq$, $qq^*$, and their conjugation/combination causes curiosity. Such behavior of hadrons indeed obeys an exact $Z_3$ symmetry that governs constituent quarks, independent of the colors. There is no necessary principle of the $B-L$ conservation as well as the $Z_3$ symmetry of quarks, since they directly result from the standard model gauge symmetry. Indeed, every interaction of the standard model separately preserves $B$ and $L$ such that $B-L$ is conserved and anomaly-free, thus quantum consistent, if right-handed neutrinos are simply imposed, while the $Z_3$ symmetry of quarks is accidentally conserved by the $SU(3)_C$ color group and never violated, because this $Z_3$ can be regarded, isomorphic to the center of the color group.  

In contrast to electric and color charges, the excess of baryons over antibaryons of the universe suggests that $B-L$ would be broken. Furthermore, $B-L$ breaking is strongly implied by compelling neutrino mass mechanisms \cite{Minkowski:1977sc,GellMann:1980vs,Yanagida:1979as,Glashow:1979nm,Mohapatra:1979ia,Mohapatra:1980yp,Lazarides:1980nt,Schechter:1980gr,Schechter:1981cv}. $B-L$ is likely to occur in the theories of left-right symmetry \cite{Davidson:1978pm,Marshak:1979fm,Mohapatra:1980qe} and grand unification \cite{Fritzsch:1974nn}, but no such traditional theories manifestly explain the existence of the accident $Z_3$ symmetry of quarks, similarly to the standard model. I point out that such hidden features of the standard model naturally arise from a $U(1)_{B-L}$ gauge symmetry. It is noted that in a period the matter parity---a residual symmetry of $B-L$ transforming trivially on normal matter---has been found usefully in supersymmetry \cite{Martin:1997ns}. I argue that there is no matter parity at all. The $Z_3$ symmetry of quarks plays the role instead in which this $Z_3$ relates to $B-L$ as the smallest and unique residual symmetry of $B-L$ itself. 

Consequently, this proposal leads to novel physical results for neutrino mass \cite{Kajita:2016cak,McDonald:2016ixn}, dark matter \cite{Jungman:1995df,Bertone:2004pz,Arcadi:2017kky}, the muon anomalous magnetic moment \cite{Abi:2021gix}, and the $W$ mass deviation \cite{CDF:2022hxs}, without necessity of any left-right symmetry, grand unification, or supersymmetry. Namely, the neutrino mass generation is induced by an inverse seesaw mechanism due to the breaking of $B-L$ by one unit. The dark matter stability is ensured by the residual $Z_3$ symmetry of $B-L$, i.e. the $Z_3$ symmetry of quarks, while the muon magnetic moment and the $W$ mass are contributed by the dark sector that contains the dark matter. 

{\it Proposal of the model}.---The full gauge symmetry is \be SU(3)_C\otimes SU(2)_L\otimes U(1)_Y\otimes U(1)_{B-L}.\label{fg}\ee Leptons and quarks transform under this symmetry as \bea l_{aL} &=& \begin{pmatrix}
\nu_{aL}\\
e_{aL}\end{pmatrix}\sim (1,2,-1/2,-1),\\
\nu_{aR} &\sim& (1,1,0,-1),\hs
e_{aR} \sim  (1,1,-1,-1),\\
q_{aL} &=& \begin{pmatrix}
u_{aL}\\
d_{aL}\end{pmatrix}\sim (3,2,1/6,1/3),\\
u_{aR} &\sim& (3,1,2/3,1/3),\hs
d_{aR} \sim  (3,1,-1/3,1/3),
\eea where the subscript $a=1,2,3$ is a family index, and the right-handed neutrinos $\nu_{aR}$ are included for $B-L$ anomaly cancelation, as usual. The gauge anomaly always vanishes if including any gauge-singlet chiral fermion (or sterile fermion), such as 
\bea N_{aL}\sim (1,1,0,0),\eea where three copies of the sterile fermion are proposed, corresponding to three families. Note that the gauge symmetry suppresses bare masses of $\nu_R \nu_R$ type, while it allows bare masses of such type for $N_L N_L$.    

The gauge symmetry breaking proceeds through the usual Higgs doublet,
\be \phi=
\begin{pmatrix}
\phi^+ \\
\phi^0\end{pmatrix}\sim (1,2,1/2,0),\ee and a scalar singlet,
\be \chi\sim (1,1,0,1),\ee that couples $N_L$ to $\nu_{R}$ through $\bar{N}_{L}\nu_R\chi$ couplings. They have vacuum expectation values (VEVs), 
\bea \langle \phi \rangle &=&
\begin{pmatrix}
0\\
v/\sqrt{2}\end{pmatrix},\hs
\langle \chi\rangle = \La/\sqrt{2}, \eea such that $\La\gg v=246$ GeV for consistency with the standard model. The scheme of symmetry breaking is 
\bc \begin{tabular}{c} $SU(3)_C\otimes SU(2)_L\otimes U(1)_Y\otimes U(1)_{B-L}$ \\
$\downarrow\La$\\
$SU(3)_C\otimes SU(2)_L\otimes U(1)_Y\otimes R$\\
$\downarrow v$\\
$SU(3)_C\otimes U(1)_Q\otimes R$ \end{tabular}\ec Here $Q=T_3+Y$ combines the weak isospin and hypercharge, as usual, whereas $R=Z_3$ is the residual symmetry of $B-L$, explicitly derived below. 

Notice that our theory does not conserve a matter parity, $M_P=(-1)^{3(B-L)+2s}$, since it is broken by $\La$, in contrast to the usual theories of $B-L$, left-right symmetry, and $SO(10)$ unification. Intriguingly, the postulate of the $B-L$ gauge symmetry and its breaking by a single $B-L$ charge, i.e. $B-L=1$, reveal important results of neutrino mass, dark matter, muon $g-2$, and $W$ mass deviation, presented in order.  

{\it Neutrino mass generation via inverse seesaw}.---The relevant Lagrangian includes
\bea \mathcal{L} &\supset& h_{ab} \bar{l}_{aL}\tilde{\phi}\nu_{bR}+f_{ab}\bar{N}_{aL}\nu_{bR}\chi -\fr 1 2 \mu_{ab} \bar{N}_{aL}N^c_{bL}+H.c.\crn 
 &\supset& -\fr 1 2 \left(\bar{\nu}_{aL}\ \bar{\nu}^c_{aR}\ \bar{N}_{aL}\right)
\!\!\begin{pmatrix} 
0 & m_{ab} & 0 \\
m_{ba} & 0 & M_{ba}\\
0 & M_{ab} & \mu_{ab}
\end{pmatrix}\!\!
\begin{pmatrix}
\nu^c_{bL} \\
\nu_{bR}\\
N^c_{bL}\end{pmatrix} +H.c.\nn\eea
Here $b=1,2,3$ is a family index as $a$ is, $\tilde{\phi}=i\sigma_2\phi^*\sim (1,2,-1/2,0)$, and a superscript $^c$ indicates charge conjugation. Additionally, the mass terms in second line are obtained by substituting the VEVs of scalars, in which $m_{ab}=-h_{ab}v/\sqrt{2}$ and $M_{ab}=-f_{ab}\La/\sqrt{2}$ are Dirac mass matrices that couple $\nu_{aL}$ to $\nu_{bR}$ and $N_{aL}$ to $\nu_{bR}$, respectively, while $\mu_{ab}$ is a Majorana mass matrix that couples $N_L$'s themselves, as given.  

$m\ll M$ is naturally imposed, since $v\ll \La$. Assuming $\mu\ll m\ll M$, the total mass matrix of neutrinos and sterile fermions takes a form of inverse seesaw \cite{Mohapatra:1986aw,Mohapatra:1986bd,Bernabeu:1987gr}. Hence, the observed neutrino mass matrix is approximately given as $\mathcal{L}\supset -\fr 1 2 \bar{\nu}_{aL} (m_\nu)_{ab}\nu^c_{bL}+H.c.$, where  
\be m_\nu\simeq m M^{T,-1}\mu M^{-1}m^T\sim (v/\La)^2\mu,\ee which is doubly suppressed by $v/\La$, in contrast to the canonical seesaw recognized in the usual $U(1)_{B-L}$ model with $B-L$ breaking by two units instead. The neutrino masses take sub-eV values suitable to observation, say $m_\nu\sim 0.1$ eV \cite{ParticleDataGroup:2020ssz}, given that $\La\sim 10$ TeV and $\mu\sim 1$ keV. Note that the mixing of $\nu_L$ with $(\nu^c_R,N_L)$ is suppressed by $m M^{-1}\ll 1$ and is thus neglected. The new fermions $\nu_R,N_L$ obtain a Dirac mass $\sim M$ at TeV scale. 

The unique property of this seesaw setup is specified as follows. Besides giving the new gauge boson mass, the $B-L$ breaking VEV, i.e. $\La$, is the largest scale in the inverse seesaw for neutrino masses. This is contrary to the conventional inverse seesaw in which $B-L$ is broken at a low scale, around keV, to induce a Majorana mass term; here, this symmetry is broken above the weak scale, giving rise to a Dirac mass term. Hence, the required smallness of such a Majorana mass, i.e. $\mu$, is not related to the $B-L$ symmetry at all. It is noted that in the limit $\mu\rightarrow 0$, our theory contains a global lepton-like symmetry, i.e. $f\rightarrow e^{i\varphi} f$ for $f=l_L,\nu_R,e_R, N_L$, which has a nature distinct from the $B-L$ gauge symmetry. Hence, the small $\mu$ is due to this symmetry protection, i.e. naturally explained by a bigger theory via a large scale or loops. Our proposal also differs from the conventional inverse seesaw in that an unreasonable Majorana mass term for $\nu_R$ is suppressed by the $B-L$ gauge symmetry, while in the conventional theory such $\nu_R$ Majorana mass arises as it has an origin identical to the $N_L$ Majorana mass.  

The $\La$ scale as given is suitable to collider constraints on the $U(1)_{B-L}$ gauge boson, called $Z'$. Indeed, the LEPII studied processes $e^+e^-\to f f^c$ for $f=\mu,\tau$ contributed by $Z'$, giving a bound $m_{Z'}/g_{B-L}>6\ \mathrm{TeV}$~\cite{ALEPH:2006bhb}. Here $g_{B-L}$ is the $U(1)_{B-L}$ coupling, and the $Z'$ mass is $m_{Z'}=g_{B-L}\La$. This translates to $\La>6$ TeV \cite{add10}. The LHC searched for dilepton signals through $pp\to ff^c$ contributed by $Z'$, supplying a bound $m_{Z'}\sim 4$ TeV for $Z'$ couplings identical to those of the $Z$ boson \cite{ATLAS:2017fih}. This converts to $\La\sim m_{Z'}/g\sim 6$ TeV, similar to the LEPII.        

{\it Residual symmetry and resultant dark sector}.---Note that $\La$ breaks only $U(1)_{B-L}$ down to $R$, whereas $v$ that breaks the electroweak symmetry obviously conserves $R$. The residual symmetry $R$ takes the form $R=e^{i\al (B-L)}$ since it is a $U(1)_{B-L}$ transformation. $R$ conserves the vacuum $\La$ if $R \La=e^{i\al (1)}\La=\La$, since $\La$ has $B-L=1$. It follows that $e^{i\al}=1$, or $\al=2\pi k $, for $k$ integer. Hence, I obtain 
$R=e^{i2\pi k(B-L)}=[w^{3(B-L)}]^k$, where $w\equiv e^{i2\pi/3}$ is the cube root of unity. The model fields transform under $R$ as in Table~\ref{tab1}, where $B-L$ is supplied for convenience in reading. It is clear that $R=1$ for every field corresponds to the smallest value of $|k|=3$, except for the identity with $k=0$. Hence, the residual symmetry $R$ is automorphic to \be Z_3=\left\{1,\mathcal{G},\mathcal{G}^2\right\},\ee where $\mathcal{G}\equiv w^{3(B-L)}$, and $\mathcal{G}^3=1$ for every field, as mentioned \cite{Ma:2015mjd}. Obviously, the residual symmetry $Z_3$ is generated by $\mathcal{G}$, called {\it matter generator}, opposite to the matter parity studied in supersymmetry. 
\begin{table}[h]
\begin{tabular}{ccccc}
\hline\hline
Field & $l$ & $q$ & $\chi$ & $\{N,\phi,A\}$ \\
\hline 
$B-L$ & $-1$ & $1/3$ & $1$ & 0 \\
$R$ & 1 & $w^k$ & 1 & 1 \\
\hline \hline
\end{tabular}
\caption[]{\label{tab1} $B-L$ charge and $R$ value of all fields, where $l$, $q$, $N$, and $A$ define every lepton (including $\nu_R$), quark, sterile fermion, and gauge boson, respectively.}
\end{table}   

\begin{table}[h]
\begin{tabular}{ccccc}
\hline\hline
Field & $l$ & $q$ & $\chi$ & $\{N,\phi,A\}$ \\
\hline 
$\mathcal{G}$ & $1$ & $w$ & 1 & $1$ \\
$Z_3$ & \underline{1} & $\underline{1}'$ & \underline{1} & \underline{1} \\
\hline \hline
\end{tabular}
\caption[]{\label{tab2} Matter generator and field representations under the residual symmetry $Z_3$.}
\end{table} 
$Z_3$ has three irreducible representations $\underline{1}$, $\underline{1}'$, and $\underline{1}''$ according to $\mathcal{G}=1$, $w$, and $w^2$, respectively. The field representations under $Z_3$ are given in Table \ref{tab2}. It is clear that every field transforms trivially under $Z_3$ with $\mathcal{G}=1$, except for quarks. Quarks are in $\underline{1}'$ with $\mathcal{G}=w$, whereas antiquarks belong to $\underline{1}''$ with $\mathcal{G}=w^2$. Hence, the hidden $Z_3$ symmetry of quarks in the standard model can be interpreted to be the residual symmetry of $B-L$. In contrast to the hidden symmetry, the residual symmetry explicitly relates to $B-L$ that would lead to dark matter with an appropriate $B-L$ value. That said, a dark field possesses a $B-L$ charge such that the matter generator is nontrivial, i.e. $\mathcal{G}=w^{3(B-L)}\neq 1$. Combined with $\mathcal{G}^3=1$ that ensures the $Z_3$ symmetry, I obtain \be B-L =\left[\begin{array}{c}
-1/3+k\\
-2/3+k'\end{array}\right. =\pm \fr 1 3,\pm \fr 2 3,\pm \fr 4 3,\pm\fr 5 3,\cdots \ee for $k,k'$ integer. This identification of dark field is independent of its spin. Additionally, the signs $\pm$ correspond to a dark field and its conjugation. Each dark field can pick up a $B-L$ charge only differing from either of the two basic charges, say $-1/3$ and $-2/3$, by an integer number, because of the cyclic property of $Z_3$. For such reasons, it is sufficient to introduce two dark fields with respect to the two basic charges, respectively; that is, a dark (Dirac) fermion and a dark vector transform under the gauge symmetry as \be F \sim (1,1,0,-1/3),\hs 
V = \begin{pmatrix}
V^0 \\
V^- \end{pmatrix} \sim (1,2,-1/2,-2/3),\nn\ee which couple to lepton doublets, \be \mathcal{L}\supset x_a \bar{l}_{aL}\ga^\mu F_L V_\mu +H.c.,\label{lbdt1}\ee in order to make the model phenomenologically viable. The detailed reason of this choice (cf. \cite{muoninteraction}) comes from the muon $g-2$, presented below. Notice that $V$ and $F$ transform under $Z_3$ as $\underline{1}'$ and $\underline{1}''$, for $\mathcal{G}=w$ and $w^2$, respectively, as given in Table \ref{tab3}.

\begin{table}[h]
\begin{tabular}{ccc}
\hline\hline
Dark-field & $V$ & $F$ \\
\hline 
$\mathcal{G}$ & $w$ & $w^2$ \\
$Z_3$ & $\underline{1}'$ & $\underline{1}''$ \\
\hline \hline
\end{tabular}
\caption[]{\label{tab3} Dark field identification according to $Z_3$.}
\end{table}  

Apart from the above couplings, $V$ and $F$ possess the Lagrangian terms \cite{note1},
\bea \mathcal{L} &\supset& \bar{F} (i \ga^\mu D_\mu - m_F) F -\fr 1 2 V^\dagger_{\mu\nu} V^{\mu \nu} +m^2_V V^\dagger_\mu V^\mu\crn
&&+ i \kappa_1 V_\mu^\dagger A^{\mu\nu} V_\nu + i \kappa_2 V_\mu^\dagger B^{\mu\nu} V_\nu+ i \kappa_3 V_\mu^\dagger C^{\mu\nu} V_\nu\crn
&&+ \al_1 (V^\dagger_\mu V^{\mu}) (V^\dagger_\nu V^{\nu}) + \al_2 (V^\dagger_\mu V^{\nu}) (V^\dagger_\nu V^{\mu}) \crn
&& +\al_3 (V^\dagger_\mu V^{\nu}) (V^{\dagger \mu} V_{\nu}) + \la_1(\chi^\dagger \chi)(V^\dagger_\mu V^{\mu})\crn
&&+\la_2(\phi^\dagger\phi) (V^\dagger_\mu V^{\mu}) + \la_3(\phi^\dagger V_{\mu})(V^{\dagger \mu}\phi),\label{lb1}
\eea where $V_{\mu\nu} \equiv D_\mu V_\nu - D_\nu V_\mu$, and $D_\mu=\pa_\mu + ig T_j A_{j\mu} + i g_Y Y B_\mu+i g_{B-L}(B-L)C_\mu$ is covariant derivative, in which $A_{\mu}$ ($A_{\mu\nu}$), $B_\mu$ ($B_{\mu\nu})$, and $C_\mu$ ($C_{\mu\nu})$ denote gauge fields (field strengths) of $SU(2)_L$, $U(1)_Y$, and $U(1)_{B-L}$, respectively. (Omitting a small kinetic mixing between two $U(1)$ gauge fields, $C$ is identical to $Z'$, while $A,B$ define $W,Z,\ga$.) This theory preserves the $Z_3$ symmetry that acts on $V,F$, in contrast to that in \cite{Saez:2018off,CarcamoHernandez:2018vdj,VanDong:2021xws}. After the symmetry breaking, the vector doublet is separated in mass, $m^2_{V^\pm}-m^2_{V^0}=\la_3v^2/2$, proportional to the weak scale, small compared to $V$ masses, $m^2_{V^\pm} = m^2_V+\la_1\La^2/2 + (\la_2+\la_3)v^2/2$ and $m^2_{V^0} = m^2_V+\la_1\La^2/2+\la_2 v^2/2$, at $\La$ scale. Dark vectors generically violate the unitarity of $S$-matrix. The unitarity condition for $\langle V V^\dagger| S | V_g V^\dagger_g\rangle$ with $V_g\in\{A,B,C\}$ constrains $\kappa_1=g$, $\kappa_2=-g_Y/2$, and $\kappa_3=-2g_{B-L}/3$, whose coefficients correspond to the gauge charges of $V$ under symmetry (\ref{fg}). This match of $\kappa_{1,2,3}$ to gauge couplings must be applied so that the theory works well up to the current energy of colliders at TeV, where the standard model is still good, in agreement with \cite{Saez:2018off}. The unitarity condition for elements like $\langle VV^\dagger| S| VV^\dagger\rangle$ and $\langle VV| S| VV\rangle$ would relate $\al_{1,2,3}$ themselves, but since $\al_{1,2,3}$ are irrelevant to the processes studied in this work, I will not refer to them further. On the other hand, interactions in (\ref{lbdt1}) also give rise to unitarity violations like $\langle V V^\dagger| S | l l^\dagger\rangle$. The unitarity is preserved, independent of $x_a$, by introducing either a new fermion or a new vector that appropriately couples to $V,l$. But, except for this role, the extra particle would not alter our results considered below, thus skipped.         
   
It is noteworthy that because $F$ and $V$ are color neutral, the lightest field of them cannot decay to colored quarks, despite the fact that both the dark field and quarks transform nontrivially under $Z_3$. Indeed, since the lightest dark field is color neutral, it cannot decay to a single quark, due to color conservation. Further, if the dark field decays to a pair of quarks, the final state must take the form $qq^c$ due to color conservation. Because $qq^c$ is trivial under $Z_3$, while the dark field is nontrivial under $Z_3$, the decay of the dark field to $qq^c$ is suppressed by $Z_3$ conservation. If the dark field decays to three kinds of quarks, the final state must take the form $qqq$ due to color conservation. But, this state is trivial under $Z_3$, while the dark field is not. Hence, the decay of the dark field to $qqq$ is suppressed by $Z_3$ conservation. Generically, if the dark field decays to a number of quarks, the final state must be composed of $qq^c$ and/or $qqq$ due to color conservation. But this final state is trivial under $Z_3$, hence suppressed by $Z_3$ conservation.   In this case, the stability of the lightest dark field is preserved by the color charge conservation, in addition to $Z_3$. This stability mechanism differs from many extensions for dark matter, including supersymmetry.

\begin{figure}[h]
\bc
\includegraphics[scale=0.6]{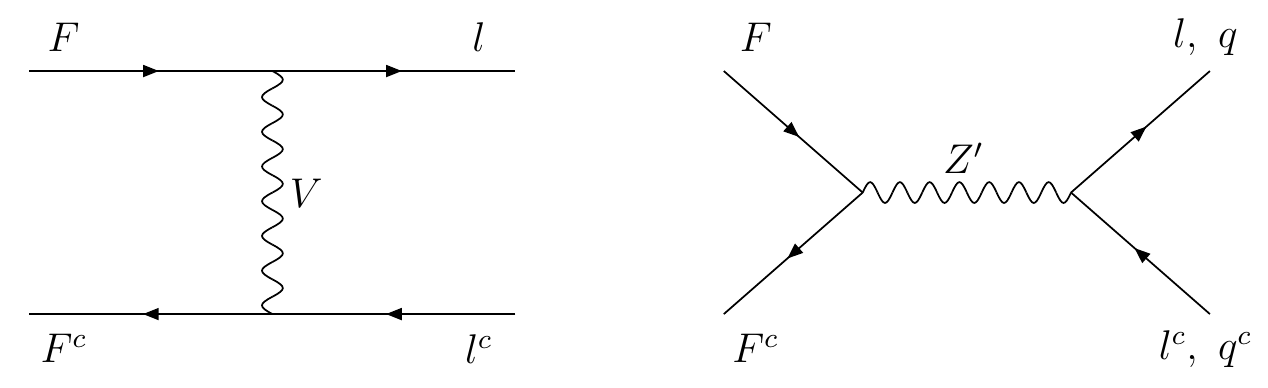}
\caption[]{\label{fig1}Dark matter annihilation to normal matter.}
\ec
\end{figure}

{\it Dark matter abundance and detection}.---There are two candidates for dark matter, $V^0$ and $F$. For the case of $V^0$, it must be the lightest of dark fields, $m_{V^0}<m_F$ and $m_{V^0}<m_{V^\pm}$ \cite{note3}. Unfortunately, this vector candidate as a complex field belongs to a weak doublet interacting with the usual $Z$ boson and is not separated in mass. The gauge interaction will induce a large scattering cross-section of $V^0$ with nuclei by $t$-channel $Z$ exchange in direct detection, which is already ruled out by experiments, analogously to the inert scalar doublet \cite{Barbieri:2006dq}. The model predicts the realistic dark matter to be a dark fermion, $F$ \cite{note2}. This fermion candidate interacts with the usual particles via $V$ and $Z'$ portals. The annihilation processes of $F$ to usual particles are described by Feynman diagrams in Figure \ref{fig1}, where we define $l=\{\nu_a,e_a\}$ for usual leptons and $q=\{u_a,d_a\}$ for usual quarks. 

As shown below for the muon $g-2$, the $V^\pm$ mass and $x_2$ coupling satisfy $|x_2|^2/4\pi m^2_{V^\pm}\sim (800\ \mathrm{GeV})^{-2}$. Hence, the $t$-channel diagram exchanged by $V$ largely contributes to the annihilation cross-section, unless $m_F$ is much smaller than $m_V$, in agreement with~\cite{VanDong:2021xws}. I also assume $m_F\ll m_{Z'}$, besides the condition $m_F\ll m_V$. Further, because of $m_{V^0}\approx m_{V^\pm}$ and $m_{Z'}=g_{B-L}\La$, the annihilation cross-section that includes both $V,Z'$ contributions as in Figure \ref{fig1} is approximated as 
\bea \langle \sigma v \rangle &\simeq& 1\ \mathrm{pb}\left(\fr{m_F}{6.5\ \mathrm{GeV}}\right)^2\left(\fr{800\ \mathrm{GeV}}{m_{V^\pm}}\right)^4 \left[\left(\fr{\sum_a |x_a|^2}{4\pi}\right)^2\right.\crn
&&\left.-\left(\fr{\sum_a |x_a|^2}{4\pi}\right)\fr{1}{6\pi}\fr{m^2_{V^\pm}}{\La^2}+\fr{37}{432\pi^2}\fr{m^4_{V^\pm}}{\La^4}\right].\eea  
This result excludes annihilation to top quarks, similarly to annihilation to right-handed neutrinos, since the dark matter is radically lighter than such fields. It is clear that $(m_{V^\pm}/\La)^2\sim 10^{-2}(|x_2|^2/4\pi)$ for $\La\sim 10$ TeV. Hence, the contributions of the $m_{V^\pm}/\La$ terms, i.e. of the $Z'$ boson, to the annihilation cross-section are small. The expression in brackets is dominated by the first term due to the contribution of $V$. Taking $\sum_a |x_a|^2/4\pi\sim 1$ in perturbative limit and $m_{V^\pm}\sim 800$ GeV similar to the muon $g-2$ below, the dark matter gets a correct abundance, i.e. $\langle \sigma v\rangle \sim 1 $ pb \cite{ParticleDataGroup:2020ssz}, if $m_F\sim 6.5$ GeV. Here I assume that there is no asymmetry in number density between a dark particle and a dark antiparticle. 

In direct detection, the dark matter $F$ scatters with quarks confined in nucleons exchanged by $Z'$, described by the effective Lagrangian, \be \mathcal{L}_{\mathrm{eff}}\supset \fr{g^2_{B-L}}{9m^2_{Z'}}(\bar{F}\ga^\mu F)(\bar{q}\ga_\mu q).\label{fae01}\ee Therefore, the scattering cross-section of $F$ on a nucleon $(p,n)$ is evaluated by 
\be \sigma_{p,n} \simeq 3.7\times 10^{-45}\left(\fr{10\ \mathrm{TeV}}{\La}\right)^4 \mathrm{cm}^2.\ee Given that $\La= 10$ TeV, the model predicts $\sigma_{p,n}\simeq 3.7\times 10^{-45}\ \mathrm{cm^2}$, in good agreement with the XENON1T experiment for dark matter mass at 6.5 GeV \cite{XENON:2017vdw,XENON:2018voc}.  

It is noted that the $Z_3$ symmetry allows only multi dark-particles produced at particle colliders. Monophoton events may be recognized at the LEPII experiment, recoiled against the missing energy carried by a pair of dark matter $F$, governed by the effective interactions \be \mathcal{L}_{\mathrm{eff}} \supset \fr{|x_1|^2}{4m^2_{V^{\pm}}}(\bar{F} \ga^\mu F)(\bar{e} \ga_\mu e)+(AA)+(VA)+(AV),\ee which are derived directly from (\ref{lbdt1}), with the aid of the Fierz identity. These vector and axial vector operators have been studied in \cite{Fox:2011fx},
leading to a bound \be m_{V^\pm}>\fr{|x_1|}{2}\times 470\ \mathrm{GeV}\sim 800\ \mathrm{GeV},\ee according to $|x_1|^2/4\pi\sim 0.92$, as expected. This mass limit agrees with the relic density and direct detection, as well as the muon $g-2$ below.

Further, monojet signals may be generated at the LHC against large missing energy carried by a $F$ pair, set by the effective interaction as in (\ref{fae01}) because the $Z'$ mediator for this process possesses a mass $m_{Z'}=g_{B-L}\La$ radically heavier than the transferred momentum ($<1$ TeV), with an appropriate $g_{B-L}$ value. Ref. \cite{Belyaev:2018pqr} has limited $g^2_{B-L}/9m^2_{Z'}=(1/3\La)^2<(1/1.1\ \mathrm{TeV})^{2}$, which is always satisfied for $\La$ at TeV. Indeed for $\La\sim 10$ TeV, the monojet signature is negligible. Additionally, since the LHC is more energetic, a pair of dark vectors each with mass about 800 GeV may be produced as $pp\to VV^\dagger$ and followed by $V,V^\dagger$ decays to stable $F$ dark matter, $V\to F^c l$ and $V^\dagger\to F l^c$, due to $Z_3$ conservation. Total cross-section is $\sigma(pp\to VV^\dagger \to FF^c ll^c)=\sigma(pp \to VV^\dagger)\times \mathrm{Br}(V\to F^cl)\times \mathrm{Br}(V^\dagger \to F l^c)$, with the aid of narrow width approximation. The cross-section $\sigma(pp\to VV^\dagger)$ is governed by $\ga,Z$ but not totally understood in this setup, since it violates unitarity similarly to the mentioned process $\langle V V^\dagger| S | l l^\dagger\rangle$, due to lack of UV completion. It is shown that relevant UV theory \cite{add12} only removes unphysical contributions arising from bad behavior of $V$ at high energy, while does not significantly modify the cross-section $\sigma(pp\to VV^\dagger)$ that comes from new fields living at UV regime $>$ 1 TeV (cf., e.g., \cite{Dion:1998pw}). Hence, $\sigma(pp\to VV^\dagger)$ is obtained by $\ga,Z$ contributions after removing the bad terms, given at quark level as $\sigma(qq^c\to VV^\dagger)\simeq (\pi \al^2/36E^{2})(1-m^2_V/E^2)^{3/2}[Q^2_q Q^2_V+Q_q Q_V v_q v_V/s^2_W c^2_W+(v^2_q+a^2_q)v^2_V/s^4_W c^4_W]$, where the energy of quark obeys $E=\fr 1 2 \sqrt{s}>m_V\gg m_Z$. I have defined $v_q = T_{3q}-2s^2_W Q_q$, $a_q = T_{3q}$, and $v_V= T_{3V}-s^2_WQ_V$, where $Q_{q,V}$ ($T_{3q,V}$) are the electric charge (weak isospin) of $q,V$, respectively. Alternatively, this cross-section can be derived, assuming the equivalence theorem $\sigma(qq^c\to VV^\dagger)\simeq \sigma (qq^c\to \Phi \Phi^\dagger)$, where $\Phi$ denotes the Goldstone boson doublet associated to $V$, which couples to $\ga,Z$ like $V$. At high energy $V$ is identical to $\Phi$ that has quantum numbers as left-handed slepton doublet, i.e. $\sigma(qq^c \to VV^\dagger)\simeq \sigma(qq^c\to \tilde{l}_L\tilde{l}^\dagger_L)$. The ATLAS \cite{ATLAS:2019lff} and CMS \cite{CMS:2020bfa} have studied a process for direct slepton production $pp\to \tilde{l}\tilde{l}^*\to ll^c \tilde{\chi}^0_1\tilde{\chi}^0_1$ assuming $\mathrm{Br}(\tilde{l}\to l\tilde{\chi}^0_1)\simeq 1$, where $\tilde{\chi}^0_1$ is the LSP dark matter, setting a bound for charged slepton mass at $700$ GeV. Given that $V$ significantly couples to $ll^c$ product, i.e. $\mathrm{Br}(V\to F^cl)\simeq 1$, the SUSY result applies to our case without change, i.e. $m_V>700$ GeV. Hence, the equivalence theorem ensures high energy behavior of $V$ as a well-studied slepton, predicting its mass limit, as expected.                                   

{\it Muon $g-2$}.---The anomalous magnetic moment of muon, $a_\mu=\fr 1 2 (g-2)_{\mu}$, in the standard model is now established, $a_{\mu}(\mathrm{SM})=116591810(43) \times 10^{-11}$ \cite{Aoyama:2020ynm}. The recent measurement of $a_\mu$ provides an exciting hint for the new physics \cite{Abi:2021gix}, in which this new result combined with the old E821 result~\cite{Bennett:2006fi} gives a deviation, \be a_{\mu}(\mathrm{Exp})-a_{\mu}(\mathrm{SM})=(251\pm 59)\times 10^{-11},\label{lb5}\ee at 4.2$\sigma$ from the standard model prediction. If this result is confirmed, many new physics approaches might be disfavored, since such deviation is larger than the electroweak contribution, say $a_\mu(\mathrm{EW})=153.6(1.0)\times 10^{-11}$, and potentially in tension with those from the electroweak precision test and current colliders.   

\begin{figure}[h]
\bc
\includegraphics[scale=0.7]{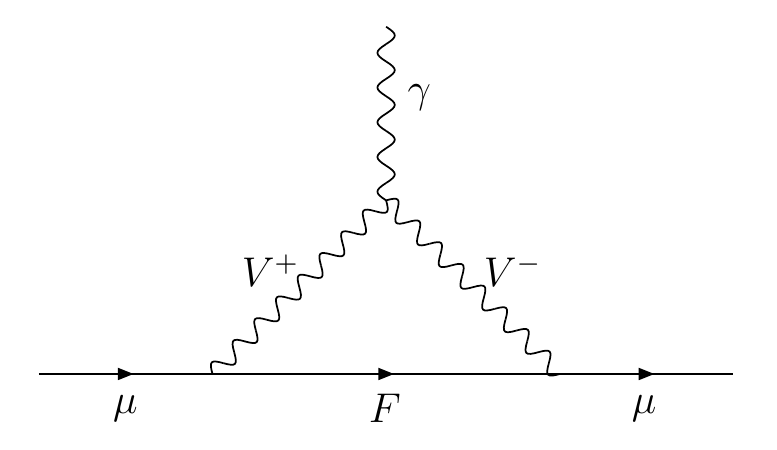}
\caption[]{\label{fig2}Dark field contribution to the muon $g-2$.}
\ec
\end{figure}      
I suggest to solve this question by a contribution from the dark sector. That said, the presence of interactions in (\ref{lbdt1}) contributes to the muon $g-2$ through a diagram given in Figure \ref{fig2}.          
Assuming $m_\mu\ll m_F,m_{V^\pm}$, I obtain 
\be \Delta a_\mu = \fr{|x_2|^2 m^2_\mu}{8\pi^2 m^2_{V^\pm}} \int^1_0 d t t \fr{t(1+t)m^2_{V^\pm}+(1-t)(1-\fr{t}{2})m^2_F}{t m^2_{V^\pm}+(1-t)m^2_F},\nn\ee where $x_2$ couples $F$ to the muon doublet of interest. The integral is of the order of 1, thus
\be \Delta a_\mu\sim 2.5\times 10^{-9}\left(\fr{ |x_2|^2}{4\pi}\right)\left(\fr{800\ \mathrm{GeV}}{m_{V^\pm}}\right)^2.\ee Compared to the muon $g-2$ deviation in (\ref{lb5}), it gives 
\be m_{V^\pm}\sim 800\sqrt{\fr{|x_2|^2}{4\pi}}\ \mathrm{GeV}.\ee 
This prediction agrees with the dark matter constraint. The $V^\pm$ field gains $m_{V^\pm}\sim 800$ GeV for $|x_2|^2/4\pi\sim 1$.  

{\it $W$ mass deviation}.---The renormalized masses of $W,Z$ in the on-shell scheme are related by $m^2_W (1-m^2_W/m^2_Z)=(\pi \al/\sqrt{2}G_F)(1+\Delta r)$, where $\Delta r=(\Delta r)^{\mathrm{SM}}+(\Delta r)^{\mathrm{NP}}$ presents quantum corrections due to the standard model and the new physics, respectively. The standard model predicts $m^{\mathrm{SM}}_W=80.357\pm 0.006\ \mathrm{GeV}$, extracted upon the precisely-measured parameters $(G_F,\al,m_Z)$ and $(\Delta r)^{\mathrm{SM}}\simeq 0.038$ \cite{Awramik:2003rn}. Given that the new physics arises as oblique contributions, one obtains $(\Delta r)^{\mathrm{NP}}=-(c^2_W/s^2_W)\Delta \rho$, where $\Delta \rho = \al(m_Z) T$ is the $\rho$-parameter deviation from the standard model related via the $T$-parameter. Recently, the CDF II collaboration has reported a novel result of $W$ mass, $m_W=80.4335\pm 0.0094\ \mathrm{GeV}$, differing from the standard model prediction at 7$\sigma$ \cite{CDF:2022hxs}. This high precision measurement of $W$ mass reveals an exciting hint for the new physics, implying $(\Delta r)^{\mathrm{NP}}\simeq -0.00489$. With $\al(m_Z)=1/128$ and $s^2_W=0.231$, it gives rise to $T\simeq 0.188$.  

In the present model, the deviation of the measured $W$ mass from the standard model expectation arises from a positive contribution of the non-degenerate vector doublet $V$ to the $T$-parameter, evaluated by
\be T = \fr{3\al^{-1}(m_Z)}{16\pi^2 v^2} \left[m^2_{V^\pm}+m^2_{V^0}-\fr{2 m^2_{V^\pm} m^2_{V^0}}{m^2_{V^\pm}-m^2_{V^0}}\ln\fr{m^2_{V^\pm}}{m^2_{V^0}}\right],\nn\ee where the coefficient 3 comes from three physical degrees of freedom of massive vectors \cite{note4}. I have included the contributions of $V$ to $W,Z$ self-energies arising from both gauge interactions of $V$ and $\kappa_{1,2,3}$ couplings furnished by the unitarity constraint. The computation in \cite{Sasaki:1992np} for $T$ in 't Hooft-Feynman gauge coincides with the above result in the unitary gauge. Notice that gauge dependence similar to the standard model $W,Z,\ga$ contributions to $T$ does not arise, since $V$ is not a gauge field \cite{Degrassi:1993kn}. Because the vector mass splitting is small, i.e. $m^2_{V^\pm}-m^2_{V^0}=\la_3v^2/2\ll m^2_{V^0}$, I further approximate 
 \be T \simeq  0.188\fr{\la^2_3}{\pi}\left(\fr{783\ \mathrm{GeV}}{m_{V^0}}\right)^2.\ee This coincides with the measured value of $W$ mass, i.e. $T\simeq 0.188$, given that \be m_{V^0}\simeq 783\sqrt{\fr{\la^2_3}{\pi}}\ \mathrm{GeV}.\ee This mass is comparable to that of the charged dark vector, if $\la_3$ is similar in size to $x_2$. 

{\it Concluding remarks}.---I have investigated a $Z_3$ symmetry of matter set by $\mathcal{G}=w^{3(B-L)}$ transformation, governing quarks as well as neutrino masses via inverse seesaw. This $Z_3$ yields two dark fields $V,F$ as potential solutions to dark matter, muon $g-2$, and $W$ mass deviation. Components of $V$ gain a mass about 800 GeV, whereas $F$ mass is at 6.5 GeV. Couplings of $V$ with leptons and Higgs boson are near perturbative limit, $|x_{2}|^2/4\pi \sim 1$, $|x_{1}|^2/4\pi \lesssim 0.92$, and $\la^2_3/\pi\sim 1$ \cite{note6}. Such $V$ also satisfies all other high energy collider bounds \cite{Saez:2018off}. The present effective theory of $V,F$ with predicted couplings reveals that the more fundamental theory may encounter either a Landau pole or a technicolor scheme above TeV \cite{note7}. 

This research is funded by NAFOSTED (Grant No. 103.01-2019.353).


\begin{thebibliography}{99}
\bibitem{add10} In the usual $B-L$ models, the $Z'$ mass is $m_{Z'}=2g_{B-L}\La$, since the $B-L$ breaking field has $B-L=2$, by contrast. This leads to $\La>3$ TeV, different from our model.
\bibitem{note1} Terms like $(\phi V)^\dagger (\phi V)$ are reducible to the given couplings (say $\la_{2,3}$), thus suppressed.
\bibitem{note3} The last condition requires $\la_3>0$, which is obviously valid if resulting from a gauge completion.
\bibitem{note2} If copies of $F$ are introduced in a UV completion, $F$ is assumed to be the lightest of such dark fermions.
\bibitem{note4} $S,U$-parameters are also modified by the $V$ doublet but suppressed by $m_W/m_V$ and $(m_W/m_V)^2$, respectively, in comparison to $T$. Hence, such contributions to the $W$ mass deviation are not significant.
\bibitem{note6} Depending on $x_{1,3}$ sizes, the value of $m_F$ derived from the density condition may be slightly smaller than 6.5 GeV, but the subsequent results remain unchanged.
\end{thebibliography}
\end{document}